\documentclass[12pt]{article}
\usepackage{graphicx}
\usepackage{lineno}
\usepackage{hyperref}

\newcommand\pubnumber{ATL-PHYS-PROC-2022-110}
\newcommand\pubdate{\today}

\def\institute{INFN Gruppo Collegato di Udine, Sezione di Trieste, Udine and ICTP, Trieste \\
Strada Costiera 11, Trieste 34151, Italy}
\newcommand\blfootnote[1]{%
  \begingroup
  \renewcommand\thefootnote{}\footnote{#1}%
  \addtocounter{footnote}{-1}%
  \endgroup
}

\def\Title#1{\begin{center} {\Large #1 } \end{center}}
\def\Author#1{\begin{center}{ \sc #1} \end{center}}
\def\Address#1{\begin{center}{ \it #1} \end{center}}

\newcommand\pubblock{\rightline{\begin{tabular}{l} \pubnumber\\
         \pubdate  \end{tabular}}}
\newenvironment{Abstract}{\begin{quotation}  }{\end{quotation}}
\newenvironment{Presented}{\begin{quotation} \begin{center} 
             PRESENTED AT\end{center}\bigskip 
      \begin{center}\begin{large}}{\end{large}\end{center} \end{quotation}}

\def\beq{\begin{equation}}
\def\eeq#1{\label{#1}\end{equation}}
\def\eeqn{\end{equation}}

\def\beqa{\begin{eqnarray}}
\def\eeqa#1{\label{#1}\end{eqnarray}}
\def\eeqan{\end{eqnarray}}

\let\bar=\overbar

\def\etal{{\it et al.}}

\def\Dslash{\not{\hbox{\kern-4pt $D$}}}
\def\dslash{\not{\hbox{\kern-2pt $\del$}}}

\def\msb{{\bar{\ssstyle M \kern -1pt S}}}

\begin{document}
\begin{titlepage}
\pubblock

\vfill
\Title{First Run 3 data/MC plots for the measurement of the top-quark pair production cross-section in pp collisions at centre-of-mass energy of 13.6 TeV with the ATLAS experiment at the LHC}
\vfill
\Author{ Giovanni Guerrieri \\ on behalf of the ATLAS Collaboration \blfootnote{\copyright $\,$ 2022 CERN for the benefit of the ATLAS Collaboration. \\
Reproduction of this article or parts of it is allowed as specified in the CC-BY-4.0 license  }}
\Address{\institute}
\vfill
\begin{Abstract}
The top quark is the heaviest known elementary particle. Its large mass, close to the scale of electroweak symmetry breaking, hints at a unique role in the Standard Model of particle physics. The study of top quark–antiquark ($t\bar{t}$) production is an important part of the physics programme of the ATLAS experiment at the CERN Large Hadron Collider (LHC). It allows quantum chromodynamics to be probed at some of the highest reachable energy scales. The $t\bar{t}$ production also forms a crucial background in many searches for physics beyond the Standard Model. Therefore, precise measurements the $t\bar{t}$ process are essential to fully exploit the discovery potential of the LHC. A comparison of data and prediction in the electron-muon final state is presented, aiming to reconstruct $t\bar{t}$ events. Data are collected in the period F1 of the LHC Run 3 at the centre-of-mass of 13.6 TeV, during the first week of August 2022, corresponding to an integrated luminosity of about 790 pb$^{-1}$.
\end{Abstract}
\vfill
\begin{Presented}
$15^\mathrm{th}$ International Workshop on Top Quark Physics\\
Durham, UK, 4--9 September, 2022
\end{Presented}
\vfill
\end{titlepage}
\def\thefootnote{\fnsymbol{footnote}}
\setcounter{footnote}{0}

\section{Introduction}
The presented plots~\cite{giova} are produced using data collected by the ATLAS~\cite{atlas} detector in $pp$ collisions at $\sqrt{s} = 13.6$~TeV at the LHC,
corresponding to an integrated luminosity of $790~$pb$^{-1}$. The selection criteria aim to reconstruct $t\bar{t}$ events.

\section{Data and MC samples}
\label{sec:datamc}

The production of $t\bar{t}$ events was modelled using the \textsc{PowhegBox} v2~\cite{Alioli:2010xd} generator at NLO with the NNPDF3.0\textsc{nlo}~\cite{Ball:2014uwa} PDF set and the $h_{\mathrm{damp}}$ parameter\footnote{The $h_{\mathrm{damp}}$ parameter is a resummation damping factor and one of the parameters that controls the matching of \textsc{PowhegBox} matrix elements to the parton shower and thus effectively regulates the high-$p_\mathrm{T}$ radiation against which the $t\bar{t}$ system recoils.} set to 1.5 $m_{\mathrm{top}}$. 
The events were interfaced to \textsc{Pythia} 8.307~\cite{Sjostrand:2014zea} to model the parton shower, hadronisation, and underlying event, with parameters set according to the A14 tune~\cite{ATL-PHYS-PUB-2014-021}. 
The $t\bar{t}$ sample is normalised to the cross-section prediction at NNLO in QCD including the resummation of NNLL soft-gluon terms calculated using \textsc{Top++} 2.0~\cite{Czakon:2011xx}. \\

The production of \(Z+\)jets was simulated with the \textsc{PowhegBox} v2 generator at NLO accuracy of the hard-scattering processes of boson production and decay in the electron, muon, and \(\tau\)-lepton channels.
The matrix element (ME) simulation is interfaced to \textsc{Pythia} 8.307 for the modelling of the parton shower, hadronisation, and underlying event. 
The CT10\textsc{nlo} PDF set~\cite{Lai:2010vv} was used for the hard-scattering processes, whereas the CTEQ6L1 PDF set~\cite{Pumplin:2002vw} was used for the parton shower.
The effect of QED final-state radiation was simulated with \textsc{Photos++} 3.64~\cite{Golonka:2005pn}.\\

The associated production of single top quarks with \(W\) bosons (\(tW\)) was modelled with the \textsc{PowhegBox} v2 generator at NLO in QCD using the five-flavour scheme and the NNPDF 3.0\textsc{nlo} set of PDFs~\cite{Ball:2014uwa}.
The events were interfaced to \textsc{Pythia} 8.307 using the A14 tune and the NNPDF 2.3\textsc{lo} set of PDFs.\\

The production of \(W+\)jets was simulated with the \textsc{Sherpa}~2.2.12~\cite{Bothmann:2019yzt} generator using NLO ME for up to two partons, and leading-order (LO) matrix elements for up to four partons calculated with the 
\textsc{OpenLoops}~\cite{Buccioni:2019sur} library.
They were matched with the \textsc{Sherpa} parton shower using 
the set of tuned parameters developed by the \textsc{Sherpa} authors.
The NNPDF 3.0\textsc{nnlo} set of PDFs was used and the samples were normalised to the NNLO prediction~\cite{Anastasiou:2003ds}.
The events are normalised to NNLO in QCD with NLO EW corrections calculated using the MATRIX \cite{Grazzini:2017mhc} software with the PDF4LHC21 PDF set.\\

Samples of diboson final states (\(VV\)) were simulated with the \textsc{Sherpa}~2.2.12 generator, including off-shell effects and Higgs boson contributions where appropriate.
Fully leptonic final states and semileptonic final states, where one boson decays leptonically and the other hadronically, were generated using
matrix elements at NLO accuracy in QCD for up to one additional parton and at LO accuracy for up to three additional parton emissions.\\

The effect of multiple interactions in the same and neighbouring bunch crossings (pile-up) was modelled by overlaying the original hard-scattering event with simulated inelastic $pp$ events generated by \textsc{Pythia}~8.307 using the NNPDF2.3 LO set of PDFs and parameter values set according to the A3 tune~\cite{ATL-PHYS-PUB-2016-017} for object with high $p_\mathrm{T}$.
For the objects with low $p_\mathrm{T}$, the EPOS~2.0.1.4~\cite{Porteboeuf:2010um} generator was used with the EPOS LHC tune. After the event generation, the ATLAS detector response is simulated
by the toolkit Geant 4~\cite{Agostinelli:2002hh} with the full simulation of the ATLAS detector.
The simulated samples are processed with the same software framework as the real data.
Additionally, dedicated correction factors are applied to the MC simulation to match the beamspot position and the number of primary vertices seen in data.

\section{Object definitions and event selection}


Electron candidates are reconstructed from electromagnetic clusters matched to particle tracks inside the ID. The candidates need to pass the \emph{TightLH} likelihood-based identification criteria~\cite{EGAM-2018-01,PERF-2017-01} with $p_\mathrm{T} > 27$~GeV and $|\eta| < 2.37$, with the transition region at $1.37 < |\eta| < 1.52$ excluded. Additionally, electron candidates need to fulfill impact parameter selection criteria: $|z_0 \sin{\theta}| < 0.5$~mm and $|d_0/\sigma(d_0)| < 5$.\\

Muon candidates are reconstructed from tracks from MS matched to tracks from ID. The candidates are required to pass \emph{Medium} identification criteria~\cite{PERF-2015-10} with ${p_\mathrm{T}>27}$~GeV and $|\eta| \le 2.47$, without accounting for hits in the New Small Wheel region. Additionally, muon candidates need to pass $|z_0 \sin{\theta}| < 0.5$~mm and ${|d_0/\sigma(d_0)| < 3}$ selection criteria. \\

Jet candidates are reconstructed from clusters of topologically connected calorimeter cells using the anti-$k_t$~\cite{Cacciari:2008gp} jet algorithm with radius parameter $R=0.4$ implemented in FastJet~\cite{Fastjet} software.
Jets are calibrated using the \emph{Particle flow} (PFlow) algorithm~\cite{PERF-2015-09} that exploits both the calorimeter as well as ID information.
After the calibration, jet candidates are required to have $p_\mathrm{T} > 30$~GeV and $|\eta| < 2.5$.

Jets containing $b$-hadrons are identified ($b$-tagged) using the DL1d algorithm.
The algorithm combines inputs from the impact parameters of displaced vertices, as well as topological properties of secondary and tertiary vertices within a jet. These are passed to a neural network that outputs three values, representing probabilities of the jet to be a light-flavour, a $c$- or a $b$-jet, which are then combined into a single discriminant. No data-driven corrections are applied to the simulation for this b-tagging discriminant, as no dedicated calibrations were available at the time of the analysis. \\

Events in the dilepton channel are required to have exactly two leptons (electrons or muons) with opposite electric charge. Only opposite flavour leptons, corresponding to the e$\mu$ channel, are considered. Events are identified by single-electron triggers, based on the legacy Level-1 calorimeter system. Furthermore, events are required to have at least one reconstructed collision vertex with two or more associated tracks with $p_\mathrm{T}>500$~MeV. The vertex with the highest $\sum p_\mathrm{T}^2$ of the associated tracks is taken as the primary vertex.\\

\begin{figure}[!htbp]
    \centering
    \begin{tabular}{cc}
      \includegraphics[width=0.48\textwidth]{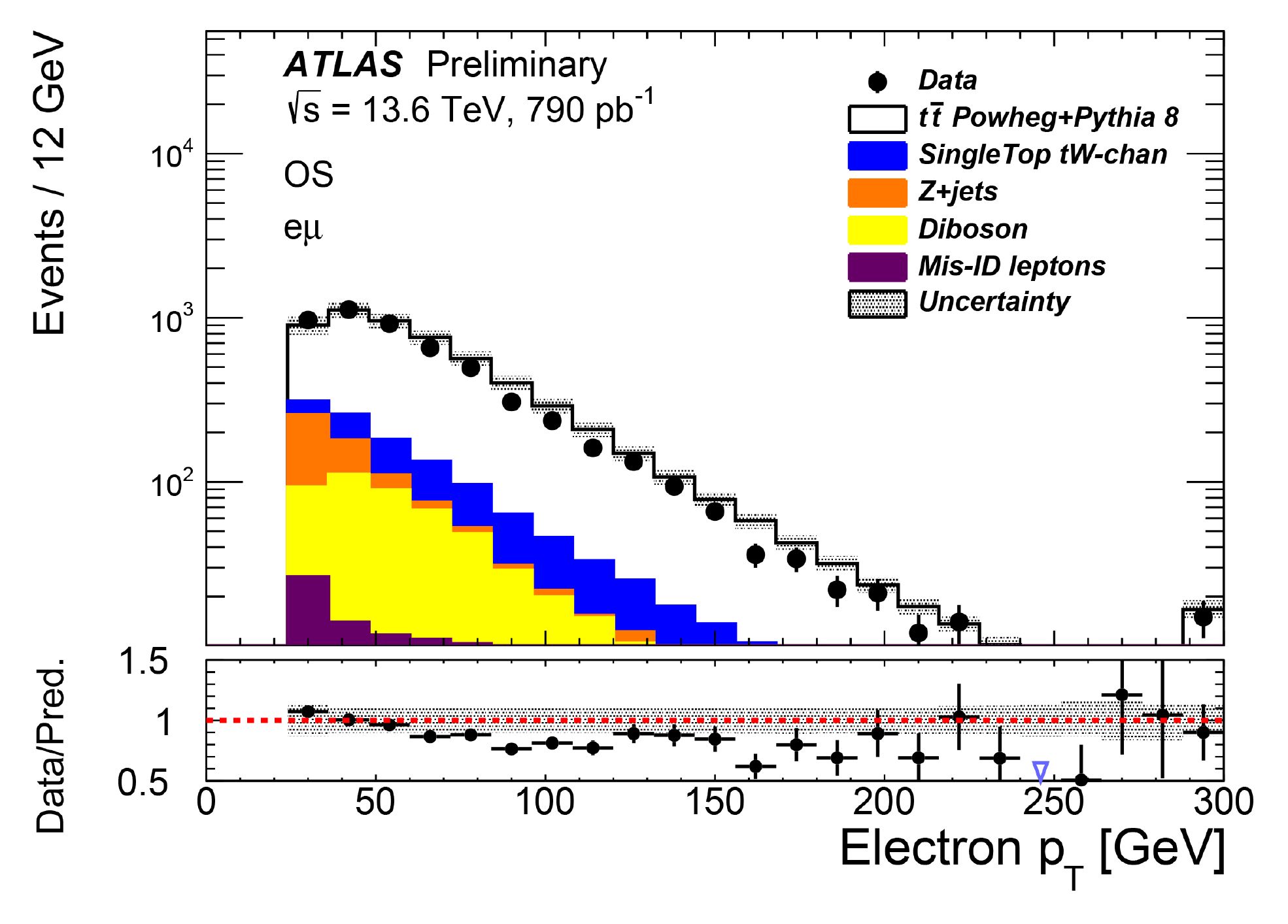} & \includegraphics[width=0.48\textwidth]{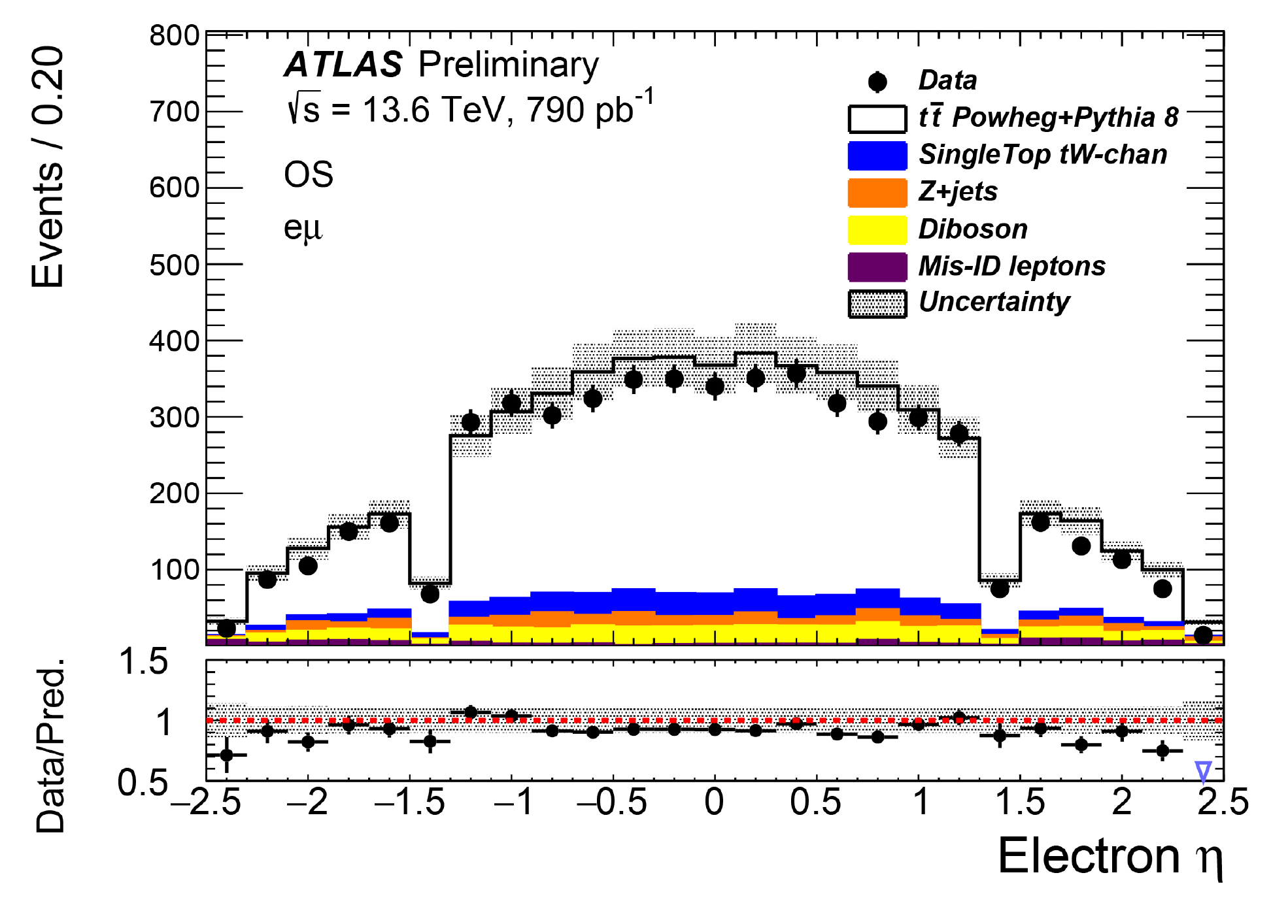}\\
      (a) & (b)\\
      \includegraphics[width=0.48\textwidth]{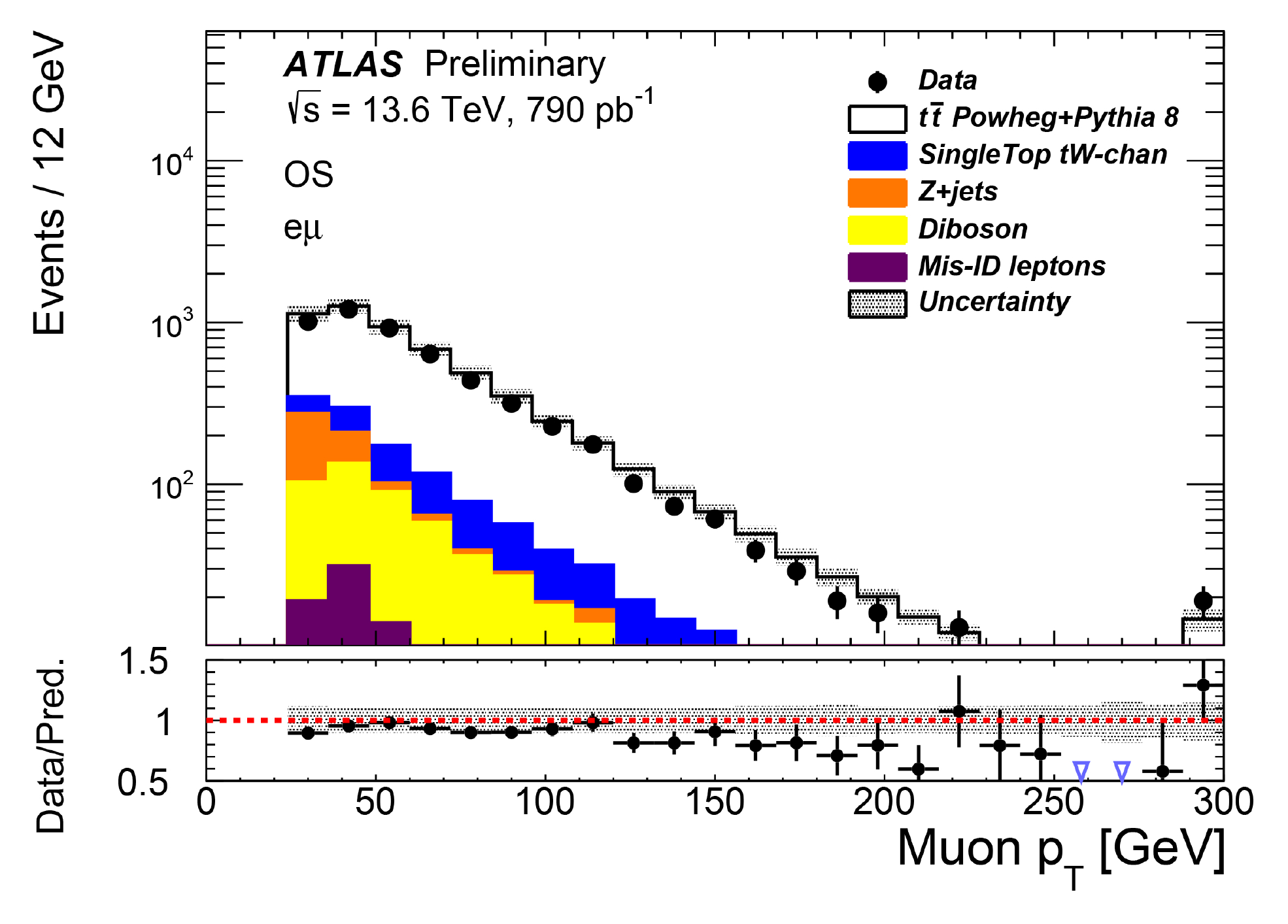} & \includegraphics[width=0.48\textwidth]{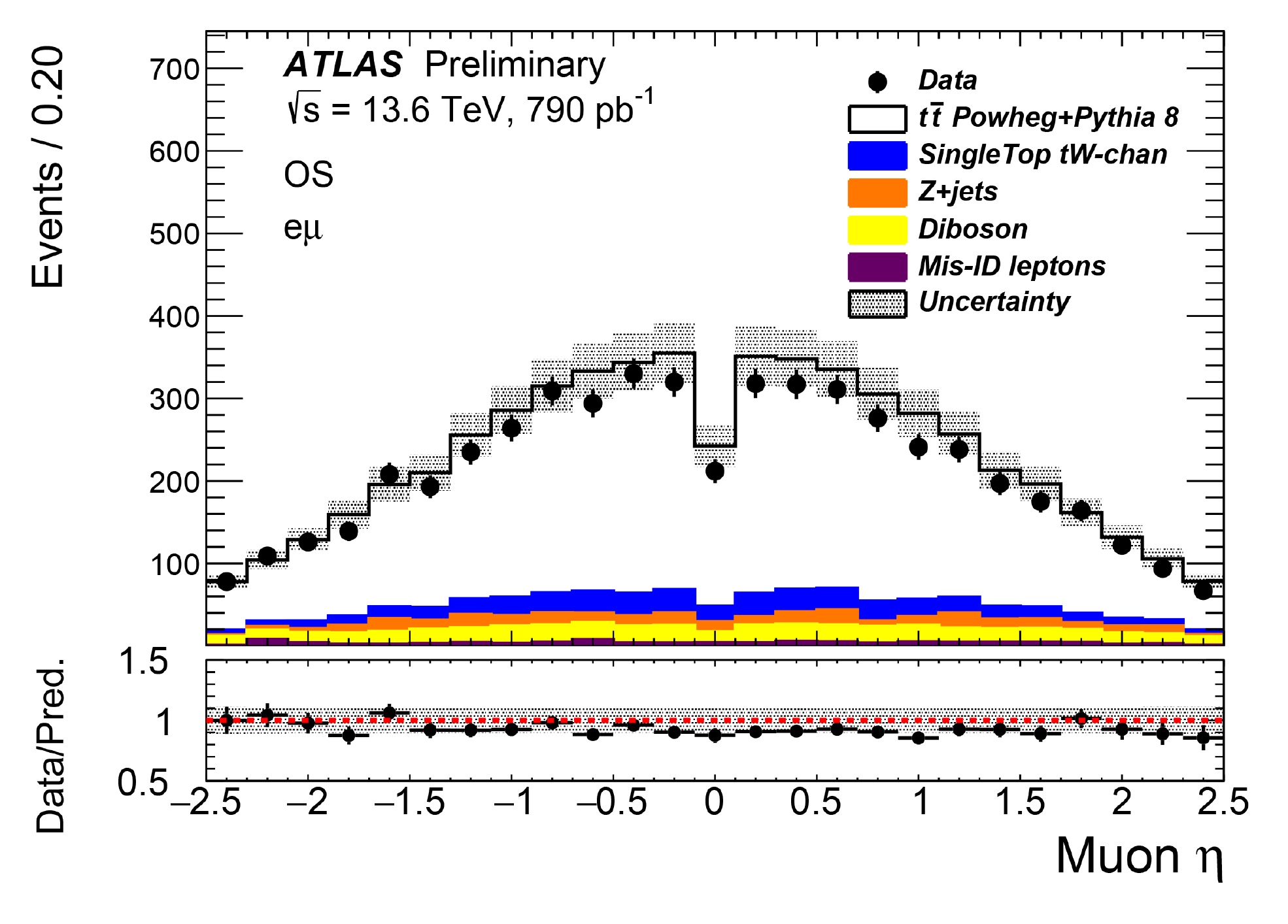}\\
      (c) & (d)\\
      \includegraphics[width=0.48\textwidth]{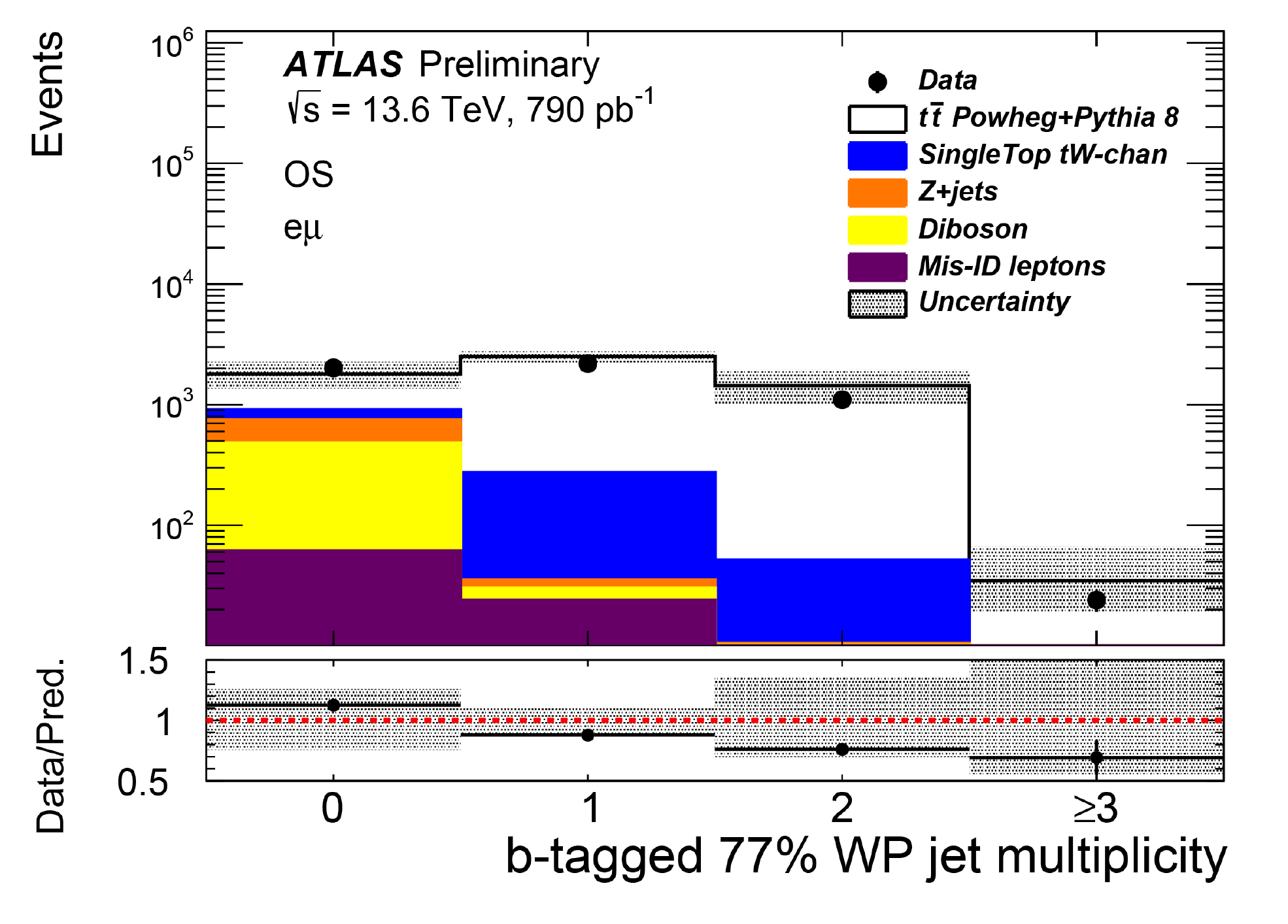} & \includegraphics[width=0.48\textwidth]{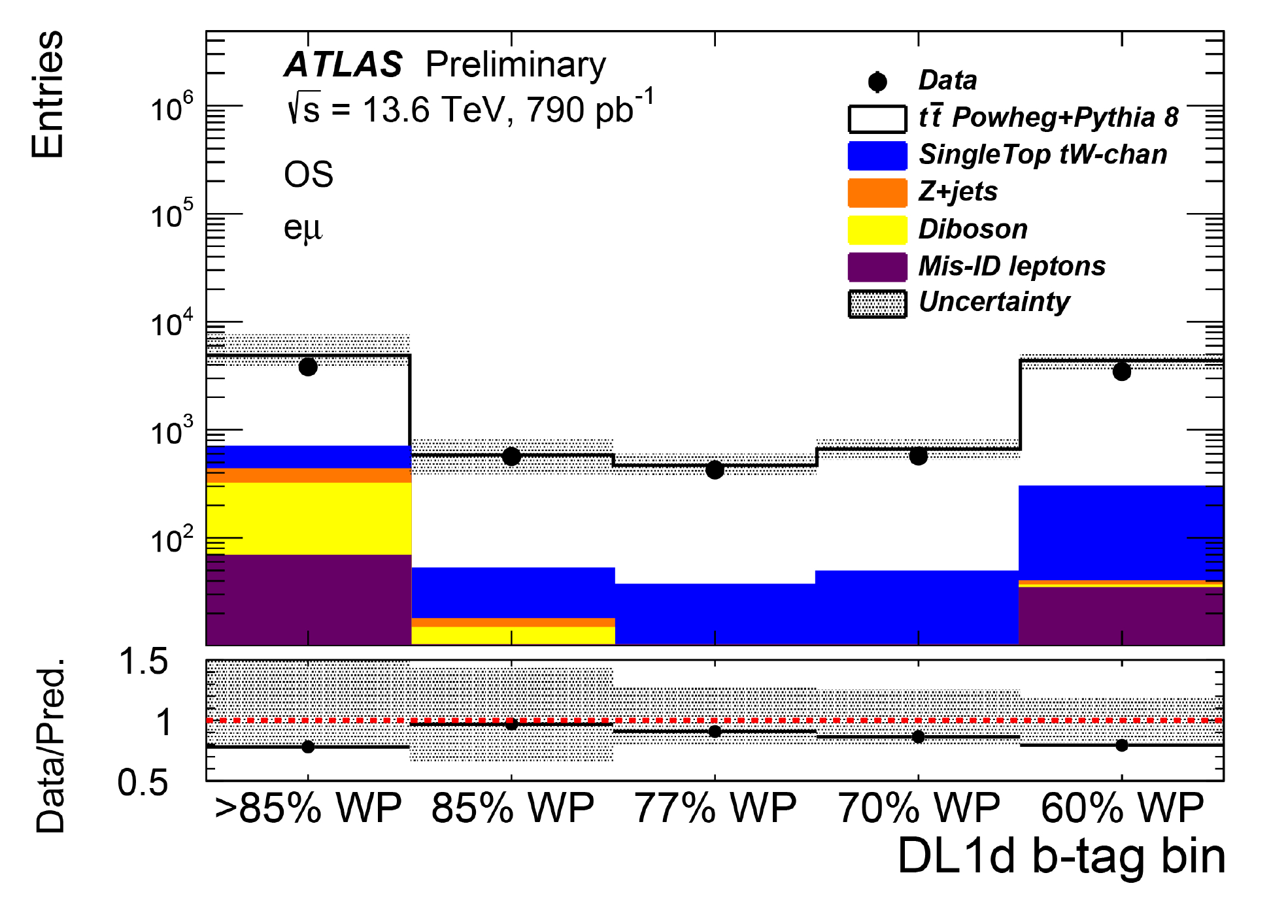} \\
      (e) & (f)\\
    \end{tabular}
  \caption{\label{fig:label}Distributions of the (a) $p_\mathrm{T}$ of the electrons; (b) $\eta$ of the electrons; (c) $p_\mathrm{T}$ of the muons; (d) $\eta$ of the muons in events with an opposite-sign $e\mu$ pair; (e) $b$-tagged jets multiplicity for the DL1d tagger at 77\% working point; (f) the amount of $b$-tagged jets for different working points of the same tagger. The data sample is compared to the expectation in the top panel of each plot, while the bottom panel shows the ratio of data over prediction. The
hashed band represents the total systematic uncertainty. The last bin in each histogram includes the overflow.~\cite{giova}\\}
\end{figure}

Figure~\ref{fig:label} shows a comparison between data and the MC prediction for the lepton $p_\mathrm{T}$ and $\eta$ distributions, as well as the jet and $b$-tagged jet multiplicity in the e$\mu$~channel. The uncertainty shown takes into account the integrated luminosity collected in the data-taking F1 period, contributing a 10\% relative uncertainty, electron energy and muon momentum corrections, together with scale and resolution inflated uncertainties obtained from the Run 2 calibrations using $Z\to ee$ and $Z\to \mu\mu$ events. Furthermore, jet-vertex-tagging and jet energy scale and resolution uncertainties are included. For flavour tagging, a conservative uncertainty, inclusive in jet~$p_\mathrm{T}$ is considered. A 10\% uncertainty for the $b$-efficiency, a 20\% uncertainty on the $c$-inefficiency and a 50\% uncertainty on the light-flavour-inefficiency is used.\\
Additionally, modelling uncertainties are applied to the $t\bar{t}$ MC process, in particular shower and hadronisation uncertainties (evaluated by considering the relative difference between \textsc{PowhegBox}~v2~+~\textsc{Pythia}~8.307 and \textsc{PowhegBox}~v2~+~\textsc{Herwig}~7.2.3~\cite{Bellm}) and the uncertainty estimated from varying the $h_{\mathrm{damp}}$ factor between $1.5\cdot m_{\mathrm{top}}$ and $3\cdot m_{\mathrm{top}}$. Simplified normalisation uncertainties are applied on background processes, specifically 50\% for $Z+$jets, diboson and lepton fakes processes, and 5.3\% for the single-top ($tW$ channel) process~\cite{run2ttbarxsec}.



\section{Summary}
The presented data vs. MC comparisons use the first data collected by the ATLAS detector in $pp$ collisions at $\sqrt{s} = 13.6$~TeV during the Run 3 of the LHC, which started in summer 2022. The plots provide a valuable input to validate the functionality of the detector and the reconstruction software which went through a number of upgrades.

\end{document}